\documentclass[aps,prl,twocolumn,showpacs,superscriptaddress]{revtex4}

\usepackage{graphicx}
\usepackage{amsmath}

\begin{document}

\title{Spin waves and magnetic exchange interactions in the spin ladder compound RbFe$_2$Se$_3$}
\author{Meng Wang}
\email{wangm@berkeley.edu}
\affiliation{Department of Physics, University of California, Berkeley, California 94720, USA }
\author{Ming Yi}
\affiliation{Department of Physics, University of California, Berkeley, California 94720, USA }
\author{Shangjian Jin}
\affiliation{School of Physics, Sun Yat-Sen University, Guangzhou 510275, China }
\author{Hongchen Jiang}
\affiliation{Stanford Institute for Materials and Energy Sciences, SLAC National Accelerator Laboratory, Menlo Park, California 94025, USA }
\author{Yu Song}
\affiliation{Department of Physics and Astronomy, Rice University, Houston, Texas 77005, USA }
\author{Huiqian Luo}
\affiliation{Beijing National Laboratory for Condensed Matter Physics, Institute of Physics, Chinese Academy of Sciences, Beijing 100190, China }
\author{A. D. Christianson}
\affiliation{Quantum Condensed Matter Division, Oak Ridge National Laboratory, Oak Ridge, Tennessee 37831, USA}
\author{C. de la Cruz}
\affiliation{Quantum Condensed Matter Division, Oak Ridge National Laboratory, Oak Ridge, Tennessee 37831, USA}
\author{E. Bourret-Courchesne}
\affiliation{Materials Science Division, Lawrence Berkeley National Laboratory, Berkeley, California 94720, USA }
\author{Dao-Xin Yao}
\affiliation{School of Physics, Sun Yat-Sen University, Guangzhou 510275, China }
\author{D. H. Lee}
\affiliation{Department of Physics, University of California, Berkeley, California 94720, USA }
\affiliation{Materials Science Division, Lawrence Berkeley National Laboratory, Berkeley, California 94720, USA }
\author{R. J. Birgeneau}
\affiliation{Department of Physics, University of California, Berkeley, California 94720, USA }
\affiliation{Materials Science Division, Lawrence Berkeley National Laboratory, Berkeley, California 94720, USA }
\affiliation{Department of Materials Science and Engineering, University of California, Berkeley, California 94720, USA }

\begin{abstract}

We report an inelastic neutron scattering study of the spin waves of the one-dimensional antiferromagnetic spin ladder compound RbFe$_2$Se$_3$.  The results reveal that the products, $SJ$'s, of the spin $S$ and the magnetic exchange interactions $J$'s along the antiferromagnetic (leg) direction and the ferromagnetic (rung) direction are comparable with those for the stripe ordered phase of the parent compounds of the iron-based superconductors. The universality of the $SJ$'s implies nearly universal spin wave dynamics and the irrelevance of the fermiology for the existence of the stripe antiferromagnetic order among various Fe-based materials.

\end{abstract}

\maketitle

Stripe antiferromagnetic (AF) order built by edge-shared Fe$X$ ($X=$ S, Se, Te, and As) tetrahedra has been characterized as one of the universal properties for the magnetic parent compounds of the iron-based superconductors\cite{Lynn2009,Dai2015}. The iron ions in the tetrahedra exhibit moments lying in-plane that order antiferromagnetically along the longer Fe-Fe bonds  and ferromagnetically along the shorter Fe-Fe bonds. Superconductivity appears when the magnetic order is suppressed by pressure or carrier doping, resulting in phase diagrams similar to those of the copper oxide based superconductors\cite{Johnston2010}. However, the origin of the stripe AF order and its relation to superconductivity are still controversial. The parent compounds of the iron-based superconductors are typically metallic with multi Fe$^{2+}$ orbitals participating near the Fermi level\cite{Graser2009}. On the one hand, in the itinerant picture the nesting between the electron and hole Fermi surfaces (FSs)  was argued to give rise to the stripe AF order\cite{Dong2008,Yang2009}. On the other hand, experimentally, insulating K$_x$Fe$_{1.5}$Se$_2$ and Rb$_x$Fe$_{1.5}$S$_2$ also exhibit the same stripe AF order in spite of the fact that all the electrons are localized\cite{Zhao2012,Wangm2014,Wang2015,Wang2015b}. Inelastic neutron scattering (INS) studies on the spin waves of the above insulating compounds have revealed comparable $SJ$'s with those of the metallic parent compounds despite the value for $S$ varies from about $1/2$ to 2; here $S$ is the total spin and the $J$'s are the magnetic exchange interactions\cite{Zhao2009,Ewings2011,Harriger2011,Zhao2014,Wang2015}. This remarkable similarity in the spin wave energy scales in the stripe AF ordered systems is yet unexplained and could suggest a universal origin of the low energy magnetic excitations. Here we test whether the above universality holds for the ladder system $A_b$Fe$_2X_3$.
\begin{figure}[b]
\includegraphics[scale=0.5]{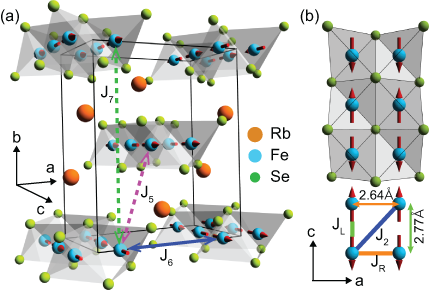}
\caption{ (a) A schmatic of the ladder structure of RbFe$_2$Se$_3$. The cuboid indicates one unit cell. (b) The 1D edge-shared FeSe tetrahedra in RbFe$_2$Se$_3$. The red arrows represent the moment directions of iron atoms. The $J_L, J_R, J_2, J_5$ and $J_6$ are the magnetic exchange interactions between the corresponding iron atoms.  }
\label{fig1}
\end{figure}

The ladder compound, $A_b$Fe$_2X_3$ ($A_b=$ Ba, K, Rb, and Cs, and $X=$ S and Se), made from the commonly observed edge-shared Fe$X$ tetrahedra with channels occupied by $A_b$, has attracted significant interest. This system hosts various magnetic structures together with some exotic characteristics. They are all insulating at ambient pressure. For BaFe$_2$Se$_3$, the irons form a block AF order with moments ($\sim2.8\mu_B$/Fe) aligned perpendicular to the leg direction, analogous to the block AF order in K$_x$Fe$_{1.6}$Se$_2$\cite{Nambu2012,Saparov2011,Bao2011}. It has been suggested that BaFe$_2$Se$_3$ is an orbital selective Mott insulator and a potential magnetic multiferroic with large ferroelectric polarization\cite{Caron2012,Dong2014,Mourigal2015}. BaFe$_2$S$_3$ exhibits stripe AF order with two identical antiferromagnetically ordered legs, with the moments ($\sim1.2\mu_B$/Fe) aligned along the rung direction\cite{Takahashi2015}. Similar to the copper-based ladder material Sr$_{0.4}$Ca$_{13.6}$Cu$_{24}$O$_{41.84}$[\onlinecite{Uehara1996}], superconductivity can be induced by pressure in  BaFe$_2$S$_3$. In this case the critical pressure is about 10 GPa\cite{Takahashi2015}. Intriguingly, for the isostructural compounds $A_a$Fe$_2$Se$_3$ ($A_a=$ K, Rb, and Cs), the moments form a similar stripe AF order along the leg direction [Fig. \ref{fig1}] with ordering moment $1.8\sim2.0\mu_B$/Fe for KFe$_2$Se$_3$ and CsFe$_2$Se$_3$\cite{Caron2012,Du2012}. A M$\mathrm{\ddot{o}}$ssbauer study suggests that the irons have an average valence close to Fe$^{2.5+}$, instead of spatially distributed Fe$^{2+}$/Fe$^{3+}$[\onlinecite{Du2012}]. The ladder system, as a quasi 1D system, may exhibit exotic characteristics. For example, both experimental and quantum Monte Carlo studies of $S=1/2$ 1D ladders have revealed a spin liquid ground state\cite{Dagotto1996}. Thus the ladder system provides a stringent test of whether the universal low energy spin wave dynamics discussed earlier also applies.

In this paper, we report INS studies on an insulating RbFe$_2$Se$_3$ powder sample. We observe two branches of spin waves associated with the stripe ordered ladders, an acoustic branch and an optical branch. From the powder averaged spectrum we are able to deduce that the acoustic branch shows a steep dispersion up to 120 meV while the optical branch is flat and centered at 205 meV. By fitting the spherically averaged experimental data to a Heisenberg Hamiltonian, we find that the spin waves can be well described by including spatially anisotropic intraladder exchange couplings ($SJ_L=70, SJ_R=-12, SJ_2=25$ meV), an interladder coupling ($SJ_6=1.5$ meV), and a single ion anisotropy term ($SJ_s=0.1$ meV), as defined in Fig. \ref{fig1}. Here, for simplicity, we have assumed a uniaxial anisotropy. The results demonstrate that the 1D stripe ordered ladders in RbFe$_2$Se$_3$ are characterized by the same universal $SJ$'s as those in the 2D stripe AF systems, again suggesting the validity of the notion of universal spin wave dynamics.

The RbFe$_2$Se$_3$ samples were grown using the Bridgman method\cite{Wangm2014}. We ground 7 g of well cleaved needle-like single crystals into powder for this experiment. Our INS experiment was carried out on the ARCS time-of-flight chopper spectrometer\cite{Abernathy2012} at the Spallation Neutron Source, Oak Ridge National laboratory (SNS, ORNL).  The powder sample was sealed in an aluminum can and loaded into a He top-loading refrigerator. The sample was measured with incident beam energies of $E_i=$ 50, 150, 250, and 450 meV at 5 K and $E_i=$ 50 meV at 300 K. The corresponding energy resolutions for these incident beams were $\Delta E=$2.2, 7.0, 13.3, and 40 meV, as determined by the full width at half maximum (FWHM) of energy cuts at $E=0$ meV. The background contributed from the sample can was not measured due to the beam time constraint and high data quality. 

Based on the zero energy transfer data obtained with $E_i=50$ meV at 5 K, we have confirmed that the nuclear and magnetic structures of RbFe$_2$Se$_3$ are consistent with those in previous reports for RbFe$_2$Se$_3$ and the stripe AF ordered ladders for KFe$_2$Se$_3$ and CsFe$_2$Se$_3$\cite{Klepp1996,Du2012,Caron2012}. The structure is in a $Cmcm$ space group (No. 63) with $a=9.42, b=11.50, c=5.54$ \AA\ at 5 K.  The nearest neighbor (NN) distance for Fe ions along the rung direction is 2.64 \AA, while the distance along the leg direction is 2.77 \AA, as shown in Fig. \ref{fig1}.

\begin{figure}[t]
\includegraphics[scale=0.5]{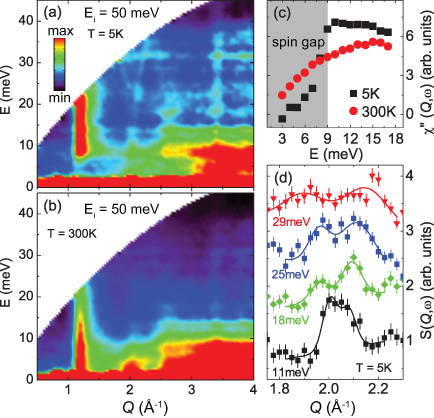}
\caption{ (a) INS spectra $S(Q,\omega)$ of RbFe$_2$Se$_3$ at 5 K and (b) 300 K with $E_i=50$ meV. The color represents intensities in arbitrary units. (c) Constant $Q$ cuts between $1.1<Q<1.3$ (\AA $^{-1})$ at 5 and 300 K, respectively. The intensities have been corrected the Bose factor. (d) Constant energy cuts at 11, 18, 25, and 29 meV at 5 K. The cuts are averaged between $E \pm1$ meV. }
\label{fig2}
\end{figure}

Figure \ref{fig2} shows INS spectra for RbFe$_2$Se$_3$ with $E_i=50$ meV at 5 and 300 K, respectively. We emphasize that because our sample is in the form of a powder the spectra are spherically averaged and therefore depend only on $|Q|$. We observe intense spin excitations with a steep dispersion relation at the momentum transfer $Q=1.21$ \AA$^{-1}$, which is consistent with the ordering wave vector $(H, K, L)=(0.5, 0.5, 1)$ of the stripe AF order. Here, $(H, K, L)$ are the Miller indices for the momentum transfer $|Q|=2\pi\sqrt{(H/a)^2+(K/b)^2+(L/c)^2}$. At 5 K, a spin gap below 10 meV at $Q=1.21$ \AA$^{-1}$, dispersive spin excitations arising from $Q=2.05$ and 2.63 \AA$^{-1}$ reaching a maximum at 35 meV, and steep excitations at $Q=3.38 $ \AA$^{-1}$ can also be seen in Fig. \ref{fig2}(a). The latter three $Q$s are consistent with the wave vectors $(H, K, L)=(2.5, 0.5, 1), (3.5, 0.5, 1)$, and (0.5, 0.5, 3). The intensities of the dispersionless excitations below 15 meV increase progressively with $Q$, consistent with the behavior of phonons. Figure \ref{fig2}(b) shows the spin excitations of the ladder structure in paramagnetic state at 300 K. The spin gap is already closed at this temperature, which is above $T_N$. However, the spin excitations are still momentum dependent, indicating that the spin correlations are preserved. The preservation of spin correlations above $T_N$ resembles the paramagnetic spin excitations of the parent compounds of the iron based superconductors, e.g., BaFe$_2$As$_2$\cite{Harriger2012a}.

To determine the spin gap and dispersion relations quantitatively, we present constant $Q$ cuts in Fig. \ref{fig2} (c) at 5 and 300 K, respectively. A background averaged between constant $Q$ cuts at $Q=0.8\pm0.1$ and $1.7\pm0.1$ \AA$^{-1}$ has been subtracted from the data and the results have then been multiplied by a Bose factor, $B(\omega, T)=1-exp(-\hbar\omega/k_BT)$, where $k_B$ is the Boltzmann constant. The intensities at 5 K show a steep step at 9 meV, demonstrating a spin gap $\Delta_s=9$ meV. In contrast, the intensities at 300 K vary smoothly across the spin gap energy. Representative constant energy cuts are presented in Fig. \ref{fig2} (d). The dispersion relation for the spin waves arising from $Q=2.05$ \AA$^{-1}$ can be extracted by fitting the cuts with two Gaussian peaks. The dispersion relation so-determined has been plotted in Fig. \ref{fig4}(a).

\begin{figure}[t]
\includegraphics[scale=0.5]{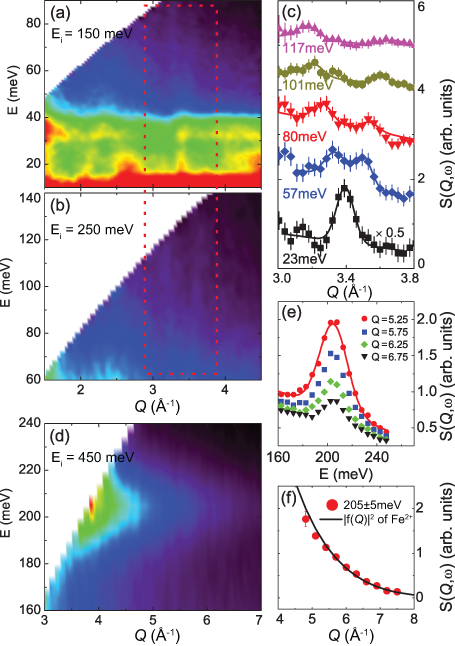}
\caption{ (a) Spin excitation spectra $S(Q, \omega)$ collected with $E_i=150$ meV and (b) $E_i=250$ meV at 5 K. The red dashed rectangular highlights a branch of dispersive spin waves. (c) Constant energy cuts at $E=23$, 57, and 80 meV averaged between $E\pm2$ meV with $E_i=150$ meV. Cuts at $E=101$ and 117 are obtained in (b) averaged between $E\pm3$ meV. The intensities for $E=23$ meV have been reduced by multiplying a factor of 0.5 for comparison. (d) $S(Q, \omega)$ collected with $E_i=450$ meV at 5 K. (e) Constant $Q$ cuts obtained in (d) at $Q=5.25$, 5.75, 6.25, and 6.75 \AA$^{-1}$ averaged between $Q\pm0.24$ \AA$^{-1}$. The solid lines in (c) and (e) are fits to Gaussian functions. (f) A plot of the intensities at $E=205\pm5$ meV after subtracting an averaged background at $E=160\pm5$ and $250\pm5$ meV. The solid line is the squared magnetic form factor of Fe$^{2+}$.  }
\label{fig3}
\end{figure}

\begin{figure}[t]
\includegraphics[scale=0.5]{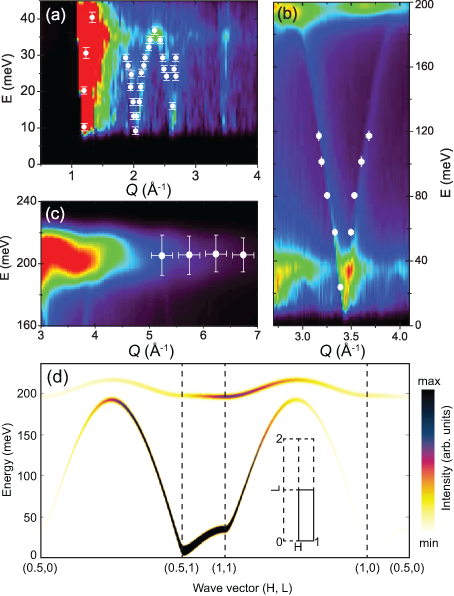}
\caption{  SpinW simulations on the $S(Q, \omega)$ with parameters described in the text. The spectrum has been convoluted an instrumental resolution of (a) 3, (b) 10, and (c) 15 meV in order to compare with the data for $E_i=50$, 250, and 450 meV, respectively. The white points are extracted from the experimental data. The bars in (c) along the vertical direction represent the FWHM. The others indicate the standard deviations of the data. (d) Simulated dispersion relation along high symmetry directions in the [H, L] 2D Brillouin zone, as shown in the inset. The color represents intensities. We convolute a constant 5 meV instrumental resolution for visualization. }
\label{fig4}
\end{figure}

In order to measure the spin excitations at higher energies, we employed incident energies of $E_i=150, 250$ and 450 meV, as shown in Fig. \ref{fig3}. In addition to the low energy spin excitations observed with $E_i=50$ meV, spin excitations stemming from $Q=3.38 $ \AA$^{-1}$ have been observed up to 120 meV, as shown in Figs. \ref{fig3}(a) and \ref{fig3}(b) with $E_i=150$ and 250 meV, respectively. The associated constant energy cuts are plotted in Fig.\ref{fig3}(c), where a clear dispersion relation can be seen. The extracted dispersion relation obtained from the fittings with two Gaussian functions has been plotted in Fig. \ref{fig4}(b).

Using incident neutrons of $E_i=250$ and 450 meV, we also observe a flat branch of excitations at $E\sim200$ meV. Figure \ref{fig3}(d) shows this flat branch with $E_i=450$ meV at 5 K. Fittings to the constant $Q$ cuts [Fig. \ref{fig3}(e)] reveal that the center of this branch is at $E=205\pm5$ meV. This flat branch resembles the optical spin waves observed in BaFe$_2$Se$_3$\cite{Mourigal2015}. To check the $Q$ dependence of the intensities, we have subtracted an averaged background integrated at $E=160\pm5$ and $250\pm5$ meV. The resulting intensities at $E=205\pm5$ meV together with a comparison with the squared magnetic form factor $|f(Q)|^2$ of Fe$^{2+}$ are presented in Fig. \ref{fig3}(f). The consistency of the observed intensities with the squared magnetic form factor $|f(Q)|^2$ confirms that the flat branch of excitations at $205\pm5$ meV is an optical branch of the spin waves in RbFe$_2$Se$_3$. We note that the magnetic form factor of Fe$^{2.5+}$ in RbFe$_2$Se$_3$ may deviate slightly from that of Fe$^{2+}$.

Having established the spin wave dispersion relations from the measured spherically averaged spectra, we proceed to extract the magnetic exchange interactions for this stripe ladder system by fitting the dispersion relations to a Heisenberg Hamiltonian, which has been widely used in the 2D stripe system\cite{Zhao2009,Ewings2011,Harriger2011,Zhao2014,Wang2015}. It can be written as
 \begin{equation} 
  \hat{H}=\frac{J_{r,r^\prime}}{2}\sum_{r,r^\prime}\bf{S}_r\cdot S_{r^\prime}-\it{J_s}\sum_r(\bf{S}_{r}^z)^2,
  \label{eq1}
 \end{equation}
where $J_{r,r^\prime}$ are the effective exchange couplings and $(r, r^\prime)$ label the iron sites, $J_s$ is the single ion Ising anisotropy term\cite{Yao2010}. (Note that technically there are in-plane and out-of-plane spin wave modes; we measure only the lowest energy gap. ) The dispersion relations can be obtained by solving Eq. \ref{eq1} using the linear spin wave approximation\cite{Yao2010}. We thus obtain analytical expressions for the spin gap $\Delta_s$, the tops of the acoustic mode both along the $H$ direction ($E_{1t}^{H}$) and $L$ direction ($E_{1t}^{L}$), and the bottom ($E_{2b}$) and top ($E_{2t}$) of the optical mode as follows:

 \begin{equation} 
 \begin{split}
 & \Delta_s=2S\sqrt{J_s(2J_L+2J_2+J_6+J_s)} , \\
 & E_{1t}^{H}=2S\sqrt{(2J_L+2J_2+J_s)(J_6+J_s)} ,\\
 & E_{1t}^{L}\approx2S(J_L+J_2) ,\\
 & E_{2b}\approx2S\sqrt{(2J_L-J_R)(2J_2-J_R)} ,\\
 & E_{2t}\approx2S(J_L-J_R+J_2) .\\
  \label{eq2}
  \end{split}
 \end{equation}
 The weak effects of $J_6$ and $J_s$ for $E_{1t}^{L}, E_{2b}$, and $E_{2t}$ have been neglected for simplification.

INS experiments on powder samples measure momentum $Q$ averaged intensities with the intensity $I(Q, \omega)\propto\frac{|f(Q)|^2}{4\pi Q^2}\int\chi^{\prime\prime}(Q,\omega)/B(\omega, T)dQ$, where $\chi^{\prime\prime}(Q,\omega)$ is the imaginary part of the dynamic susceptibility\cite{Mutka1991}. Presumably, the observed spin excitations stemming from $Q=1.21, 2.05, 2.63$ and 3.38 \AA$^{-1}$ correspond to the wave vectors of $(H, K, L)=(0.5, 0.5, 1), (2.5, 0.5, 1), (3.5, 0.5, 1)$ and $(0.5, 0.5, 3)$. The dispersion relations between 2.05 and 2.63 \AA\ and around $Q=3.38$ \AA\ can be attributed to the dispersions along the $H$ and $L$ directions, respectively. The extracted parameters, e.g., $\Delta_s=9$ meV, $E_{1t}^{H}=35$ meV, the dispersion relation along the $[0.5, 0.5, L]$ direction below 120 meV, and the center of the optical branch $E_{op}=205\pm5$ meV, set strong constraints on the exchange interactions. By comparing with the experimental data, we find a set of parameters ($SJ_L=70\pm5, SJ_R=-12\pm2, SJ_2=25\pm5, SJ_6=1.5\pm0.2$ and $SJ_s=0.1\pm0.01$ meV) that perfectly fits these constrains. There should exist other weak out-of-ladder plane exchange couplings, e.g., $J_5, J_7$, that give rise to the three dimensional magnetic order, as defined in Fig. \ref{fig1}(a). However, determining $SJ_5$ and $SJ_7$ would require observation of the dispersion relation along the $K$ direction, which is beyond the scope of our measurements on powdered samples.

Previous studies on an $S=1/2$ spin liquid ladder system with an even number of legs revealed a finite spin gap (known as `a Haldane gap'), corresponding to the lowest $S=1$ excitations\cite{Haldane1983,Dagotto1996}. To check whether such quantum mechanical effects contribute to this observed 9 meV gap, we carried out numerical calculations using the density-matrix renormalization-group (DMRG) method\cite{White1992,Jiang2010} based on the determined $SJ_L, SJ_R$, and $SJ_2$ for both $S=1$ and 2 cases. The results reveal that the values for both Haldane gaps would be 3.03 meV for $S=1$ and 0.41 meV for $S=2$. The spin $S$ for RbFe$_2$Se$_3$ is likely to be the same as that for Rb$_x$Fe$_{1.5}$S$_2$\cite{Wang2015}, namely $S=2$, since both systems have the same insulating ground state and similar Fe$X$ tetrahedra. Thus, the observed 9 meV gap would still be dominated by the single ion anisotropy gap in RbFe$_2$Se$_3$.

To check the consistency of the fitted spectra and the experimental data directly, we have also used the SpinW program, which employs classical Monte Carlo simulations and linear spin wave theory, to simulate the spherically averaged spin wave spectra\cite{Toth2015}. The simulated spectra together with the experimentally determined dispersion relations are presented in Figs. \ref{fig4}(a)-\ref{fig4}(c). Instrumental resolutions of 3, 10, and 15 meV have been convoluted with the calculated profiles for comparisons with the $E_i=50, 250$ and 450 meV data. The simulations match well with the experimental data. In addition, we have also plotted the dispersion relations along high symmetry directions in the [H, L] plane in Fig. \ref{fig4} (d). These results would be useful for comparison with spin waves measured on single crystal samples.

\begin{table}[t]
\caption{The magnetic exchange couplings and spin states in the stripe AF order of iron pnictides and chalcogenides\cite{Zhao2009,Ewings2011,Harriger2011,Zhao2014,Wang2015}. The $J_{1a}$ and $J_{1b}$ are equivalent to the $J_L$ and $J_R$ for RbFe$_2$Se$_3$, repectively.}
\begin{tabular}{lccccc}
\hline \hline
Compounds               & $SJ_{1a}$   & $SJ_{1b}$  & $SJ_2$ (meV) & $S$   & $M(\mu_B)$   \\ \hline
CaFe$_2$As$_2$   & $50\pm10$      & $-6\pm5$      & $19\pm4$   & 0.5  & $0.80$ \\
BaFe$_2$As$_2$   & $59\pm2$      & $-9\pm2$      & $14\pm1$   & 0.5  & $0.87$ \\
SrFe$_2$As$_2$   & $39\pm2$      & $-5\pm5$      & $27\pm1$   & $0.69$  & $0.94$ \\
K$_2$Fe$_3$Se$_4$         & $38\pm7$     & $-11\pm5$     & $19\pm2$     &   & 2.8\\
Rb$_2$Fe$_3$S$_4$ & $42\pm5$      & $-20\pm2$    & $17\pm2$     & 2   & $\sim2.8$    \\ 
RbFe$_2$Se$_3$ & $70\pm5$      & $-12\pm2$    & $25\pm5$     &   & $1.8\sim2.0$    \\ \hline \hline
\end{tabular}
\label{table:t1}
\end{table}

We list the parameters describing the spin wave dynamics of the 2D and 1D stripe AF ordered iron pnictide and iron chalcogenide materials in Table \ref{table:t1}. The parent compounds of the iron pnictide superconductors $T_M$Fe$_2$As$_2$ ($T_M=$ Ca, Sr, and Ba) are bad metals consisting of 2D AF iron planes\cite{Rotter2008,Huang2008}. Both itinerant electrons and local moments exist, resulting in a small moment ($<1\mu_B$/Fe) and a spin of $S\sim0.5$\cite{Zhao2009}. In contrast, the 2D stripe AF ordered $A_x$Fe$_{1.5}X_2$ ($A=$ K, Rb; $X=$ Se, S) with a rhombic iron vacancy order are insulating with purely localized electrons stabilized in a high spin configuration $S=2$ with a large moment size ($\sim2.8\mu_B$/Fe)\cite{Zhao2012,Wangm2014,Wang2015,Wang2015b}. However, RbFe$_2$Se$_3$ consists of quasi-one-dimensional iron ladders, which can be viewed as cuts from a 2D stripe AF order along the AF ordering direction. It is also an insulator with a medium moment size of $1.8\sim2.0\mu_B$/Fe\cite{Du2012,Caron2012}. The strong anisotropic $J_L$ and $J_R$ could originate from structural orthorhombicity and possible orbital orderings, similar as that in $A_x$Fe$_{1.5}X_2$\cite{Zhao2012,Wang2015}. The three types of systems exhibit distinct ground states, nuclear structures, electronic structures, spin configurations, and Fermi surface topologies. Intriguingly, the products of $SJ$'s for the AF direction ($SJ_{1a}$ and $SJ_{L}$), the ferromagnetic (FM) ordering direction ($SJ_{1b}$ and $SJ_{R}$), and the diagonal direction ($SJ_2$) are comparable. They all exhibit FM order and FM exchange interactions between the shorter NN Fe-Fe ions, and AF order and AF exchange interactions between the longer NN Fe-Fe ions\cite{Zhao2009,Ewings2011,Harriger2011,Zhao2014,Wang2015}. The results reveal that the stripe AF order is robust given the edge-shared Fe$X$ tetrahedra and that the Fermi surface topology is not crucial for the formation of the stripe AF order. Moreover, as suggested by our data, all these systems exhibit similar `universal' spin wave energy scales!

In summary, through INS measurements on powder samples we have successfully mapped out two branches of the spin waves spectra associated with the stripe AF order in the 1D iron ladder compound RbFe$_2$Se$_3$. We have thereby determined the magnetic exchange interactions by fitting the data to a linearized spin wave prediction of the Heisenberg Hamiltonian. The results reveal that the 1D system has comparable products of $SJ$'s with the 2D stripe systems. The universality of the $SJ$'s suggest that the spin wave dynamics is nearly universal for all stripe AF ordered iron-based materials. 

This work was supported by the Office of Science, Office of Basic Energy Sciences, Materials Sciences and Engineering Division, of the U.S. Department of Energy under Contract No. DE-AC02-05-CH11231 within the Quantum Materials Program (KC2202) and the Office of Basic Energy Sciences U.S. DOE Grant No. DE-AC03-76SF008. The research at Sun Yat-Sen University was supported by NBRPC-2012CB821400, NSFC-11275279, NSFC-11574404, and NSFG-2015A030313176. HCJ was supported by the Department of Energy, Office of Science, Basic Energy Sciences, Materials Sciences and Engineering Division, under Contract DE-AC02-76SF00515. H. Luo wish to thank the support from NSFC and MOST of China. The experiment at Oak Ridge National Laboratory's Spallation Neutron Source was sponsored by the Scientific User Facilities Division, Office of Basic Energy Sciences, U.S. Department of Energy.



\bibliography{mengbib}

\end{document}